\documentclass[a4paper]{article}

\usepackage{INTERSPEECH2022}
\usepackage{float}
\usepackage{amsmath,graphicx,bbm,mathrsfs,caption,subfig}
\usepackage{booktabs,diagbox,multirow}
\usepackage{dutchcal,authblk}
\usepackage{longtable}
\usepackage{authblk}
\usepackage{hyperref}
\usepackage{wrapfig}
\usepackage{enumitem}

\title{Self-supervised Context-aware Style Representation for Expressive Speech Synthesis}
\name{Yihan Wu$^{1,\dagger}$\thanks{$^\dagger$Work done during an internship at Microsoft.}, Xi Wang$^{2}$, Shaofei Zhang$^{2}$, Lei He$^{2}$, Ruihua Song$^{1,*}$\thanks{$^*$ Corresponding author:  songruihua\_bloon@outlook.com.}, Jian-Yun Nie$^{3}$}
\address{$^1$ Gaoling School of Artificial Intelligence, Renmin University of China, China\\
$^2$ Microsoft Azure Speech, China \\
$^3$ Université de Montréal, China}

\email{yihanwu@ruc.edu.cn, \{xwang, shazh, helei\}@microsoft.com, nie@iro.umontreal.ca}

\begin{document}

\maketitle
\begin{abstract}
Expressive speech synthesis, like audiobook synthesis, is still challenging for style representation learning and prediction. 
Deriving from reference audio or predicting style tags from text requires a huge amount of labeled data, which is costly to acquire and difficult to define and annotate accurately.  
In this paper, we propose a novel framework for learning style representation from abundant plain text in a self-supervised manner. It leverages an emotion lexicon and uses contrastive learning and deep clustering. We further integrate the style representation as a conditioned embedding in a multi-style Transformer TTS.
Comparing with multi-style TTS by predicting style tags trained on the same dataset but with human annotations, our  method achieves improved  results according to subjective evaluations on both in-domain and out-of-domain test sets in audiobook speech. Moreover, with implicit context-aware style representation, the emotion transition of synthesized audio in a long paragraph appears more natural. The audio samples are available on the demo website. ~\footnote{~\href{https://wyh2000.github.io/InterSpeech2022/}{https://wyh2000.github.io/InterSpeech2022/}}

\end{abstract}

\section{Introduction}
Although TTS models can synthesize clean and high-quality natural speeches, it still suffers from the issue of over-smoothing prosody pattern in some complex scenarios, as in audiobook synthesis. One of the reasons is the difficulty of modeling high-level characteristics such as emotions and context variations, which impact the overall prosody and speaking style.
Being different with the low-level acoustic characteristics such as duration, pitch and energy, modeling high-level characteristics is more challenging and crucial in these complex scenarios~\cite{pmlr-v80-skerry-ryan18a}. 

There are two general approaches to deal with such tasks: unsupervised joint training and supervised label conditioning.
The unsupervised approach models styles based on joint training with both reference audio and text content ~\cite{pmlr-v80-skerry-ryan18a, akuzawa2018expressive, 9023186, pmlr-v80-wang18h, stanton2018predictGST}. By constructing an implicit style representation space in an unsupervised way, it infers a style representation from either the reference audio or the predicted style by the joint training process. However, the joint training framework faces two challenges: 1) content information leaks into style encoder; 2) requiring a large number of audio and content pairs. Many recent studies in this area focus on these two issues ~\cite{hu2020unsupervised,sun2020generating,ma2018neural}. In real applications, the supervised learning is more widely adopted by leveraging explicit labels as auxiliary information to guide multi-style TTS  ~\cite{lee2017emotional,kim2021expressive}. It does not require reference audio, but the definition of styles, which could be subjective. Predicting style tags also requires a large amount of annotated data. Moreover, a simple discrete tag cannot fully reflect the nuance in speech styles.

To address these problems, instead of modeling styles through reference audios or explicit tags, we propose a novel framework which learns the style representation from plain text in a self-supervised manner and integrates it into an end-to-end conditioned TTS model. First, we employ contrastive learning to pre-train style embedding by distinguishing between similar and dissimilar utterances. To this end, we create a similar utterance by replacing an emotional word by a similar one, determined using an emotion lexicon. With the emotionally similar utterance as positive sample, all other dissimilar utterances in the randomly sampled minibatch are treated as negatives. Then training samples in style embedding space are clustered by minimizing deep clustering loss, reconstruction loss and contrastive loss together. We learn the style representation from a large amount of unlabeled plain text data and construct a text sentiment embedding space to guide the generation of multi-style expressive audio in speech synthesis. Using it as a pre-training of style information, we can get rid of the dependence of matched audio and content. Our work has three main contributions:
\begin{figure*}
  \centering
  \includegraphics[width=0.8\linewidth]{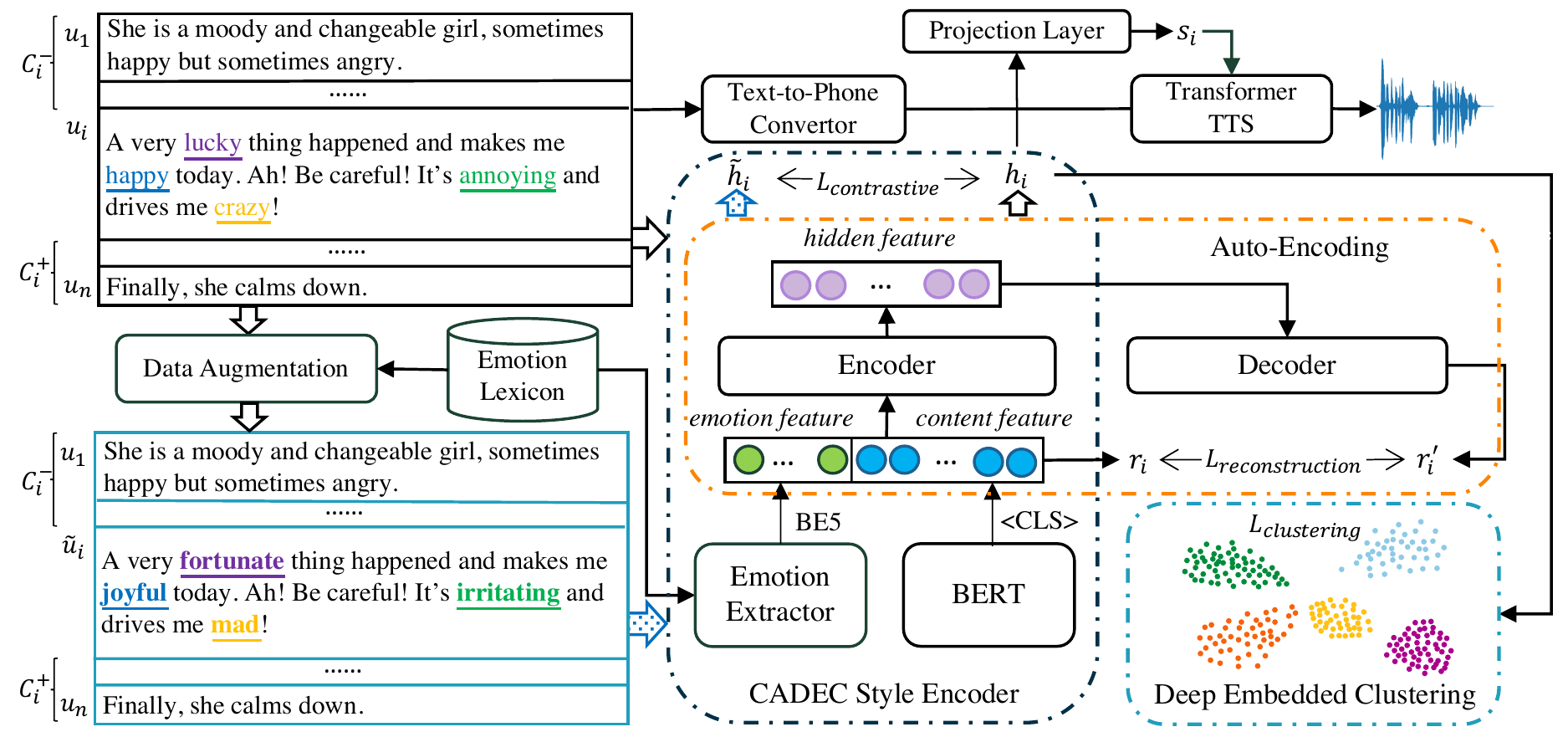}
  \caption{Overview of our proposed framework} 
  \label{framework} 
  \vspace{-1.0em}
\end{figure*}
\begin{itemize}[leftmargin=*]
\item[$\bullet$] 
We propose a novel framework for modeling style representation from unlabeled texts and incorporate it into a style-based TTS model, without  reference audio or explicit style labels. 
\item[$\bullet$] We propose a novel two-stage style representation learning method  combining deep embedded clustering with contrastive learning based on  data augmented via an emotion lexicon. 
\item[$\bullet$] We demonstrate that with the same labeled text corpus and audiobook corpus, our  speech synthesis outperforms the baseline, especially in naturalness of emotion transition in long audio generation.
\end{itemize}

\section{Related Work}
\label{sec:format}
Our work is  related to contrastive learning and deep clustering.

Contrastive learning is one of the most efficient ways to extract useful information from massive unlabeled data. Chen et al.~\cite{pmlr-v119-chen20j} propose a simple contrastive framework called simCLR which learns visual representations by maximizing the similarity between a similar pair. Furthermore, SimSiam~\cite{chen2021exploring} explores simple siamese network to learn meaningful representation without negative sample pairs. Besides, contrastive learning is also widely used in NLP and speech areas ~\cite{48253,bose2018adversarial,9383605,9413711}. Compared to the existing work, we propose a new data augmentation strategy based on emotion strength and apply simCLR to our scenario.

Clustering is an effective method of analyzing data without category annotations. Deep embedded clustering (DEC)~\cite{pmlr-v48-xieb16} maps the observed data to a low-dimensional space and optimizes KL divergence as clustering objective. Improved deep embedded clustering (IDEC)~\cite{guo2017improved} adds reconstruction loss to maintain original data structure. Many extension models have been proposed, leading to high performance in various tasks~\cite{do2021clustering, park2021improving, zhan2020online}. Combining with contrastive learning,
Supporting Clustering with Contrastive Learning (SCCL) ~\cite{zhang2021supporting} achieves state-of-the-art results on short text clustering task by jointly optimizing clustering loss and contrastive loss. Inspired by these studies, we combine contrastive learning and deep clustering to improve expressive TTS systems for the first time.

\section{Our Approach}
\subsection{Problem Formulation and System Overview}
\label{ssec:subhead}
Assume a text dataset $\mathcal{D} = \left\{{U_i}\right\}_{i=1}^D$ where $U_i=\{{C_i^-}, u_i, {C_i^+}\}$ represents text utterance $u_i$ and its context. $ C_i^- = \{u_{i-m},..., u_{i-1}\}$ is the preceding utterances of $u_i$ and $C_i^+=\{u_{i+1},...,u_{i+m}\}$ is the following utterances of $u_i$. 
Our goal is to learn a style encoding model $s_i=g(U_i)$ from $\mathcal{D}$ which can generate a context-aware style representation for utterance $u_i$. This style model will be applied to a TTS system to improve the expressiveness of speech synthesis.

Figure~\ref{framework} shows our proposed framework. 
First, we construct positive pairs of $u_i$ and its augmented sample $\tilde{u}_i$ by replacing the words having the strongest emotion arousal by their synonyms. Based on the data, we design a style encoder and pre-train it via contrastive learning. Second, to optimize the global distribution of our style representation, we further enhance the improved Deep Embedded Clustering method~\cite{guo2017improved} with contrastive learning to train our style encoder further. Through the two stages, we learn $g(U_i)$ and denote the generated representation as Context-aware Augmented Deep Embedded Clustering (CADEC) style. Finally, we feed it into Transformer TTS as conditioning embedding to generate expressive audio by applying appropriate style to text.

\begin{table*}[h]
\footnotesize
\centering
\caption{subjective evaluation of proposed model and baseline on \textbf{TTS-evaluation set}}
\label{subjective}
\begin{tabular}{l|l|lll|ll|ll}
\toprule  
\multicolumn{2}{c|}{Metrics} &\multicolumn{3}{c|}{MOS} &\multicolumn{2}{c|}{CMOS} & \multicolumn{2}{c}{Paragraph CMOS}\\ 
\midrule
\multicolumn{2}{c|}{Settings} & Recording & Baseline & Our Model& Baseline & Our Model & Baseline & Our Model \\ 
\midrule
\multicolumn{2}{c|}{in-domain} & $4.35 \pm 0.01$ & $4.26 \pm 0.07$ & \textbf{$4.34 \pm 0.06$} & $ 0 $ & \bm {$+ 0.22$} & $-$ & $-$\\
\midrule  
\multirow{2}{*}{out-of-domain} & Female & $4.37\pm0.1$& {$4.21 \pm 0.07$}& \textbf{$4.28 \pm 0.06$}  & $0$ & \bm{$+ 0.03$} &$ 0 $& \bm{ $+0.22$} \\
& Male & 4.35$\pm$0.13 & {4.1$\pm$0.1}& \textbf{$4.18 \pm 0.09$} & $0$ & \bm{$+ 0.05$} & $0$ & \bm{$+ 0.23$}\\
\bottomrule 
\end{tabular}
\end{table*}

\subsection{Stage 1: Contrastive Learning with Data Augmentation}
\label{ssec:subhead}
An utterance could be expressed by human in various styles. The appropriate style of utterance $u_i$ is highly related to context, its semantic content and conveyed emotion. We propose taking $u_i$ and its context together, i.e. $U_i$, as input and combine both content feature and emotion feature to model the best-fit style. 

We employ a pretrained BERT~\cite{devlin2019bert} as backbone to extract content features, and an extra emotion lexicon~\cite{buechel2020learning} to extract emotion features. The emotion lexicon starts from a manually annotated English source emotion lexicon. Combining emotion mapping, machine translation, and embedding-based lexicon expansion, the monolingual lexicons for 91 languages with more than two million entries for each language are created. The lexicon provides word-level emotion features including VAD (valance, arousal, dominance) on 1-to-9 scales and BE5 (joy, anger, sadness, fear, disgust) on 1-to-5 scales. Then, we extract our initial style embedding $r_i$ by:
\begin{equation}
\setlength\abovedisplayskip{3pt}
\setlength\belowdisplayskip{3pt}
    r_i=b(U_i)\oplus\frac{1}{M}\sum_{j=1}^{M}{e(w_j)}
\end{equation}
where $\oplus$ denotes a concatenation operator, $b(U_i)$ is the output [CLS] embedding by inputting $U_i$ into BERT, $M$ is the total number of words in $U_i$ and $w_j$ is $j$-th word in $U_i$ while $e(w_j)$ denotes its normalized BE5 feature which is a 5-dimensional vector. 

Then we add a fully connected multilayer perceptron (MLP) as encoder to map the initial embedding into hidden features, which are our output style embedding: 
\begin{equation}
\setlength\abovedisplayskip{3pt}
\setlength\belowdisplayskip{3pt}
    h_i=MLP(r_i)\label{mlp}
\end{equation}
We propose augmenting data and using contrastive learning to pre-train the parameters of encoder. 

To augment $u_i$ to the utterance $\tilde{u}_i$ that would have similar speech style, we first split $u_i$ into shorter segments not longer than a fixed length, e.g., 10 in our experiments. Then we look up the emotion lexicon to get emotion arousal for each word in a segment and select top $k$\%, e.g., 20\%, to be replaced by their WordNet synonyms~\cite{morris2020textattack}. Take the utterance $u_i$ in Figure~\ref{framework} as an example. We split it into two segments, and select ``lucky'' and ``happy'' in the first segment and ``annoying'' and ``crazy'' in the second segment. We then replace them with their synonyms to compose $\tilde{u}_i$. The aim of splitting a long sentence into segments is to extract emotional words from different segments, thereby avoiding focusing on the dominant emotional words from some segment only.
For example, although ``fortunate'' has higher arousal than ``annoying'' in the whole sentence of 20 words, we avoid choosing it for the whole sentence by the segment-based selection. This makes our concentrated emotional words more evenly distributed to ensure the expressiveness of the whole sentence.

As for contrastive learning, from a large training dataset $\mathcal{D}$, we randomly sample a minibatch data $\mathcal{B}=\left\{{U_i}\right\}_{i=1}^N $, and generate its augmented data $\mathcal{\tilde{B}}=\{\tilde{U}_i\}_{i=1}^N $, where $\tilde{U}_i=\{C_i^-,\tilde{u}_i,C_i^+\}$. $ U_i $ and $ \tilde{U}_i $ are treated as positive pairs while the other $N-1$ pairs $\{<U_i,\tilde{U}_k>\}_{i\neq k}$ are all negative examples in one minibatch. To maximize the agreement between texts with similar emotions and disagreement between texts with different emotions, following simCLR~\cite{pmlr-v119-chen20j}, we calculate the sample-wise contrastive loss by
\begin{equation}
\setlength\abovedisplayskip{3pt}
\setlength\belowdisplayskip{3pt}
    \mathcal{l}_c^i=-log{\frac{exp(cos(h_i,\tilde{h_i})/{\tau})}{\sum_{k=1}^{N}\mathbbm 1_{k\neq i}exp(cos(h_i,\tilde{h_k})/{\tau})}}\label{equ2}
\end{equation}
Here $\tau$ is the temperature parameter and $ \mathbbm 1_{k\neq i} $ is the indicator function. The contrastive loss for a minibatch is computed by averaging over all instances in $\mathcal{B}$ and its augmented data $\tilde{\mathcal{B}}$:
\begin{equation}
\setlength\abovedisplayskip{3pt}
\setlength\belowdisplayskip{3pt}
    \mathcal{L}_{contrastive}=\frac{1}{N}{\sum_{i}^{N}\mathcal{l}_c^i}\label{equ3}
\end{equation}

As the significant overlap of initial representation, this stage proves useful as the start of Stage 2 in our experiments.

\subsection{Stage 2: Deep Embedded Clustering with Autoencoder}
\label{ssec:subhead}
To optimize the global distribution of style representations, we apply deep embedded clustering with autoencoder to train the CADEC style encoder further.
The number of clusters K is a prior and each cluster is represented by its centroid $\mu_k $.
Clustering loss is defined as
\begin{equation}
\setlength\abovedisplayskip{3pt}
\setlength\belowdisplayskip{3pt}
    \mathcal{L}_{clustering}=KL(P \| Q)=\sum_{i}\sum_{k}p_{ik}log\frac{p_{ik}}{q_{ik}}
\end{equation}
where $P$ is the target distribution of $Q$.
Following ~\cite{van2008visualizing}, we apply the Student's \textit{t}-distribution to compute the probability of assigning $h_i$ to the $k^{th}$ cluster $ q_{ik}$.
\begin{equation}
    q_{ik} = \frac{{(1+{\|{h_{i}- \mu_{k}}\|}_{2}^{2}/\alpha)}^{-\frac{\alpha+1}{2}}}{\sum_{k^{'}=1}^{K} (1+{\|{h_{i}- \mu_{k^{'}}}\|}_{2}^{2}/\alpha)^{-\frac{\alpha+1}{2}}}
\end{equation}
where $\alpha$ denotes the degree of freedom of the Student's t-distribution. In this work, we set $\alpha=1 $.
The target distribution $p_{ik}$ is
\begin{equation}
\setlength\abovedisplayskip{3pt}
\setlength\belowdisplayskip{3pt}
   p_{ik} = \frac{q_{ik}^{2}/\sum_{i}{q_{ik}}}{\sum_{k^{'}}(q_{ik^{'}}^{2}/\sum_{i}{q_{ik^{'}}})}
\end{equation}

As this kind of clustering distorts the original space of representation and weakens the representation ability of implicit feature ~\cite{guo2017improved}, we add an autoencoder structure with reconstruction loss as in Equation ~\ref{reconstructionloss}, in order to preserve local structure of feature space and avoid corruption:
\begin{equation}
\setlength\abovedisplayskip{3pt}
\setlength\belowdisplayskip{3pt}
    \mathcal{L}_{reconstruction}=\sum_{i=1}^{N}{{\|r_i-r_{i}^{'}\|}_2^2}\label{reconstructionloss}
\end{equation}

Therefore, the objective of Stage 2 is defined as 
\begin{equation}
\setlength\abovedisplayskip{3pt}
\setlength\belowdisplayskip{3pt}
    \mathcal{L_{total}}=\mathcal{L}_{contrastive}+\beta \mathcal{L}_{clustering}+\gamma \mathcal{L}_{reconstruction}
\end{equation}

We tune all trainable parameters to optimize $\mathcal{L_{total}}$ and obtain the final style encoder of our CADEC style.

\subsection{TTS Stage: Transformer TTS with Style Representation}
\label{ssec:subhead}
Transformer TTS~\cite{li2019neural} leverages Transformer-based encoder-attention-decoder architecture. It performs well in synthesizing high-quality audio in fast speed. We extend Transformer TTS architecture by conditioning on CADEC style embedding generated from our proposed style encoder, with phoneme sequence and its style representation as inputs, as shown in Figure~\ref{framework}. Due to space limitation, we refer the reader to the paper~\cite{li2019neural} for more details about conditioned Transformer TTS.


\begin{figure}[t]
  \centering
  \includegraphics[width=.8\linewidth]{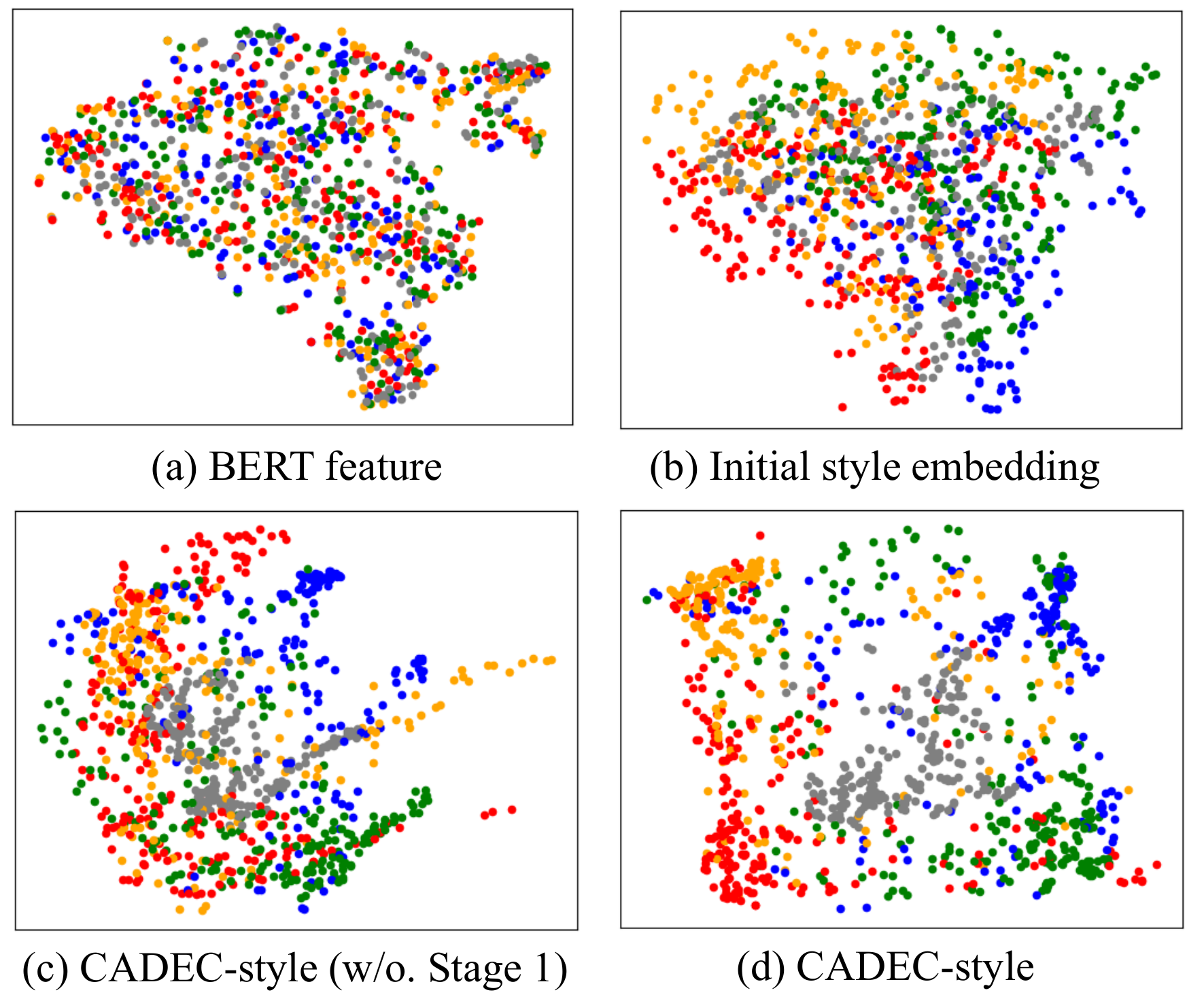}
  \caption{Visualization of the embedding space learned from different models. Each color indicates a ground truth emotion category. (a) represents the original BERT [CLS] embedding without fine-tuning. (b) represents initial style embedding which combines BERT embedding with BE5. (c) represents style embedding learned from our proposed model without Stage 1. (d) represents our learned CADEC style embedding.} 
  \label{t-SNE} 
  \vspace{-1.5em}
\end{figure}

\section{Experiments}
\label{sec:typestyle}
\subsection{Setup}
To train and evaluate CADEC style encoder and TTS model, we use two datasets respectively, a plain text dataset for CADEC style encoder and an audiobook dataset with (text, audio) pairs for Transformer TTS model.
\begin{itemize}[leftmargin=*]
\item \textbf{Text-training set and text-evaluation set.}
For CADEC encoder, we collect a plain text dataset from e-books. It contains 1.7M conversation utterances with contexts. We split the dataset into three sets, 1,696,000 conversation utterances as \textbf{text-training set}, 2000 utterances as
\textbf{in-domain text-evaluation set} for validation and 2000 utterances as \textbf{out-of-domain text-evaluation set} for testing. ~\footnote{It should be mentioned that for supervised baseline approach training and accuracy evaluation, all conversations utterances are annotated into five emotion categories (calm, joy, serious, fear and depressed) by crowdsourcing, but we do not use these annotations during the CADEC style encoder training. }
\item[$\bullet$] \textbf{TTS-training set and TTS-evaluation set.}
For TTS model, we apply an internal audio corpus as \textbf{TTS-training set},  read by a female speaker and a male speaker in Chinese. It is a complete 14 hours of expressive storytelling book, in which a portion of about 4 hours is conversation audio. Following most previous TTS work, we set in-domain test set from same audiobook and out-of-domain test set respectively to evaluate our model’s performance in different domains. We random select 10,981 utterances as training set, 200 utterances as \textbf{in-domain TTS-evaluation set} and 500 utterances from other books as \textbf{out-of-domain TTS-evaluation set}. Specifically, to evaluate voice quality of different genders, \textbf{out-of-domain TTS-evaluation set} contains both female and male speakers.
\end{itemize}

We use a supervised system that performed well in industrial as baseline. It contains two components: style prediction model and multi-style TTS model. The style prediction model (denoted as BERT-style) fine-tunes a pretrained BERT~\footnote{\href{https://huggingface.co/bert-base-chinese}{https://huggingface.co/bert-base-chinese}} with a downstream emotion classification task by leveraging annotated plain text dataset in a supervised manner. The annotated style tag is also applied when training baseline's multi-style based Transformer TTS model.

In our proposed self-supervised style learning TTS framework, we use the same BERT backbone with content length of 256 as input in the same way as baseline BERT-style. In Stage 1 and Stage 2, we adopt Adam optimizer with mini-batch size of 32 and the initial learning rate as 1e-6. We firstly train 1,000 epoches with the contrastive loss to fine-tune BERT and autoencoder's encoder in Stage 1 and continue to optimize all modules by minimizing $ \mathcal{L_{total}} $ in Stage 2 until convergence. By Grid Search, we set convergence threshold as 0.1\%, and both the coefficient $\beta$ and $\gamma $ of the clustering loss as 0.5. 
For multi-style TTS system training, narrative utterance's style embedding is zero-initialized in our proposed method. Both the baseline and our proposed method use similar settings including the acoustic feature as 80-dimensional log mel-spectrogram, window shift 12.5 ms. A MelGAN vocoder~\cite{kumar2019melgan} of 24kHz is used for both.

\begin{table}[tb]
\footnotesize
\centering
\setlength{\tabcolsep}{5mm}{
\caption{classification results of five emotion categories on \textbf{text-evaluation set}}
\label{acc}
\begin{tabular}{lc}
    \toprule
    Models & Overall accuracy ($\%$) \\
    \hline 
    BERT-style & 57.63\\
    CADEC-style & 40.20\\
    CADEC-style w/o. Stage 1  & 31.15\\
    CADEC-style w/o. context  & 32.82\\
    \bottomrule 
\end{tabular}}
\vspace{-1.0em}
\end{table}

\subsection{Subjective Evaluation of TTS}
\label{ssec:subhead}
To evaluate the effectiveness of style embedding in TTS, we conduct subjective listening tests on Microsoft UHRS crowdsourcing platform. Participants are required to focus on speech style and expressiveness in all tests. For sentence-level speech, both MOS (Mean of Opinion Score) and CMOS (Comparative MOS) are conducted on \textbf{in-domain TTS-evaluation set} and \textbf{out-of-domain TTS-evaluation set}. Each audio is judged by at least 10 participants. As shown in Table~\ref{subjective}, compared with the baseline, our model achieves better voice quality in terms of both MOS and CMOS, and importantly without any explicit annotation labels. The improvement in different datasets also demonstrates the effectiveness and robustness of our proposed method to cross-domain datasets of different genders.

In the audiobook scenario, a paragraph CMOS is used to evaluate the style expressiveness of multiple continuous utterances, as a paragraph or a session, in a conversation with narrative context. The judges are asked to rate audio considering context, which reflects the coherence and appropriateness of the audios' expressiveness. We randomly select 20 paragraphs from \textbf{out-of-domain TTS-evaluation set} for each speaker with synthesized speech exceeding 30 seconds. The result shows significant preference over baseline, i.e. 0.22 for female and 0.23 for male speakers respectively. It verifies our assumption that style representation would be a continuous embedding other than a simple tag especially in a long context. Our  model can express much more appropriate and diverse styles according to context, and achieves natural emotion transition between sentences. 

\subsection{Analysis of Style Embedding}
\label{ssec:subhead}
Compared with BERT-style that requires predefined categories, CADEC style embedding achieves relatively lower accuracy in emotion classification tasks (Table~\ref{acc}). In further analyzing text and human-annotated emotion labels, we find that five discrete labels can not model complex context-aware emotion representation, which leads to low classifier accuracy.
However, as our proposed model performs prominently better than the baseline in TTS evaluation experiments, it demonstrates that continuous embedding is more suitable for TTS tasks. Meanwhile, removing Stage 1 leads to a significant drop in accuracy (Table~\ref{acc}) and implicit embedding space overlap (Figure~\ref{t-SNE}(c)). This ablation study shows the importance of Stage 1 and further demonstrates that our proposed CADEC style embedding is effective in learning styles in addition to content. Moreover, removing text's context reduces the overall accuracy by 7.18\% which demonstrates the importance of context in style modeling. More experiment results and audio samples could refer to \href{https://wyh2000.github.io/InterSpeech2022/}{https://wyh2000.github.io/InterSpeech2022/}


\section{Conclusion and Future Work}
\label{sec:majhead}
In this work, we present a novel framework for self-supervised
context-aware style encoding from unlabeled text for text-to-speech.
Our proposed model has several advantages. 1) As a pre-trained style representation model from text, it can learn from massive unannotated data without requiring the corresponding audio.
2) by combining context-aware information and modeling style information in a continuous feature space, it can achieve natural expressiveness and emotion transition in long paragraph. In the future, we will  enhance text style representation with a larger amount of text corpus for better accuracy and robustness. Besides, we will further explore mapping style embedding from text into acoustic feature space more robustly and guiding synthesized speech style in a flexible and controllable manner.

\bibliographystyle{IEEEtran}

\bibliography{mybib}

\end{document}